\documentclass[12pt]{article}
\pdfoutput=1

\usepackage{formatting}
\usepackage{hyperref}
\usepackage{wrapfig}
\usepackage{amsbsy}

\begin{document} 
\begin{titlepage}
\begin{center}
\phantom{ }

{\bf \large{Constraining large $N$ CFTs\\
from holographic entanglement}}
\vskip 0.5cm
{Guglielmo Grimaldi}
\vskip -.10cm

\begin{abstract}
We prove that for general $d$-dimensional conformal field theories, the validity of a family of holographic entropy inequalities at large $N$ for any thermal state implies that the theory's vacuum free energy dominates for every
\begin{equation*}
    \beta > \pi \frac{\text{Vol}(\mathbb{S}^{d-1})}{\text{Vol}(\mathbb{S}^{d-2})} > \beta_{\text{HP}}\,,
\end{equation*}
where $\beta_{\text{HP}}$ is the Hawking-Page phase transition threshold. Our argument makes no assumption on the theory's spectrum, so we interpret it as a purely entropic understanding of spectrum sparsity in holographic theories. As an application of this result, we show that it predicts a phase transition in the mutual information for two disjoint ball regions in any dimension.

Specializing to $d = 3$, we also prove that validity of the same family of inequalities at large $N$ for the vacuum implies an upper bound on the ratio $C_T/F_0$ between the stress-tensor two-point function coefficient and the sphere free energy 
\begin{equation*}
    \frac{C_T}{F_0} \leq \frac{3}{\pi^3} \approx 0.09675\, .
\end{equation*}
We use both constraints to give support and refine the conjecture that holographic entropy inequalities are a property of large $N$ field theories, by checking them against weakly coupled $3d$ ABJM theory and $4d$ $\mathcal{N} = 4$ SYM. 

\end{abstract}
\end{center}

\end{titlepage}

\setcounter{tocdepth}{3}

\hrule 
{\parskip = .2\baselineskip \tableofcontents}

 \vskip .5cm
 \hrule

\section{Introduction}
Entanglement entropy has a long history of teaching us new aspects of quantum field theory, in part because the entanglement entropy of a region contains information about universal data of the theory. Universal properties of entanglement entropy can then constrain such QFT data. For example, given three disjoint spatial regions $A,B,C$ on a Cauchy slice $\Sigma$ of the QFT, we have the strong subadditivity (SSA) inequality
\begin{equation}\label{eq:ssa}
    S(AB) + S(BC) \geq S(B) + S(ABC)\,,
\end{equation}
where $S(X) = -\Tr \rho_X \log \rho_X$ is the von Neumann entropy associated to the reduced density matrix $\rho_X = \Tr_{\bar{X}} \rho$. The inequality \eqref{eq:ssa} implies many fundamental theorems of quantum field theory. In $d = 2$ it implies that the central charge $c$ decreases monotonically along RG flow, and in $d = 3$ it implies the same for the sphere's free energy $F_0 = -\log Z[\mathbb{S}^3]$ which appears as the universal finite non-local term in the entanglement entropy of a disk. These results are known as the $c$-theorem \cite{Casini:2006es, Casini:2004bw} (see also \cite{Zamolodchikov:1986gt} for the original proof that doesn't use SSA and \cite{Kudler-Flam:2026abf} for a recent proof using relative entropy) and the $F$-theorem \cite{Casini:2012ei}. An analogous result is also known for $d = 4$, the $a$-theorem \cite{Casini:2017vbe}. Since these universal terms can be understood as characterizing the number of degrees of freedom for the theory, these irreversibility theorems formalize the intuition that one loses information when coarse graining a physical system.

Restricting to a particular class of theories, entanglement entropy and its properties can be constrained further. This is especially true for conformal field theories (CFTs) with a holographic dual. In the classical limit (large $N$, strong coupling) of AdS/CFT, where the bulk is well-described by Einstein gravity, the entanglement entropy of a boundary spatial region $A$ is realized geometrically via the Ryu-Takayanagi formula
\begin{equation}
    S(A) = \min_{\gamma \sim A} \frac{\text{area}(\gamma)}{4G_{\text{N}}}\,,
\end{equation}
as the area of the minimal surface homologous to $A$ \cite{Ryu:06b20v, Ryu:2006ef, Hubeny:2007xt}. 

Crucially, this formula satisfies an infinite tower of linear inequalities beyond the SSA inequality \eqref{eq:ssa}, in the following sense. Given $\mathsf{N}$ boundary spatial regions, one can draw all the RT surfaces associated with both fundamental and composite boundary regions for a total of $2^{\mathsf{N}}-1$ entropies. These can be collected into an \emph{entropy vector}
\begin{equation}
    \vec{\mathsf{S}} = (S(A_1), S(A_2), \cdots, S(A_1 A_2\dots A_{\mathsf{N}}))\,,
\end{equation}
which lives in \emph{entropy space} $\mathbb{R}^{2^{\mathsf{N}-1}}$. For a given $\mathsf{N}$, the set of physically realizable entropy vectors forms a convex polyhedral cone $H_{\mathsf{N}}$ known as the holographic entropy cone (HEC), whose facets are defined by a set of linear inequalities called holographic entropy inequalities (HEIs) \cite{Bao:2015bfa}. The HEIs characterize the entanglement structure of classical holographic states, and they are necessary constraints for any quantum state to admit a classical dual geometry.  

The first among these holographic-specific constraints is known as monogamy of mutual information (MMI) \cite{Hayden:2011ag} which appears already at $\mathsf{N} = 3$,  and reads
\begin{equation}\label{eq:mmi}
    S(AB) + S(BC) + S(AC) \geq S(A) + S(B) + S(C) + S(ABC).
\end{equation}
We stress that MMI, as well as all the other HEIs, are \emph{not} obeyed by general quantum states. Familiar examples that violate MMI are the 4-party GHZ state, the ground state of a massive scalar field, and holographic states incorporating quantum $1/N$ corrections. 

Beyond MMI, the systematic study of HEIs was initiated by the authors of \cite{Bao:2015bfa} who rewrote the problem of proving the validity of an HEI as a graph-theory question. It is now understood that the set of HEIs is equivalent to the set of constraints obeyed by the min-cut function on weighted graphs. The full set of inequalities for $\mathsf{N} = 3, 4,5$ is known \cite{HernandezCuenca:2019wgh}, and for $\mathsf{N} = 6$ only partly with tens of thousands of HEIs found \cite{Hernandez-Cuenca:2023iqh, hecdatacurrent}. Understanding the underlying structure of these constraints for general $\mathsf{N}$ remains an exciting open question. However, a few highly symmetric families of HEIs are known for arbitrary large $\mathsf{N}$ \cite{Czech:2024rco}. One particularly simple example is the \emph{dihedral family}, when $\mathsf{N} = 2k + 1$ is odd we have 
\begin{equation}\label{eq:cyclic}
    \sum_{i = 1}^{2k + 1} S(A_iA_{i+1}\dots A_{i+k}) - \sum_{i = 1}^{2k + 1} S(A_iA_{i+1}\dots A_{i+k-1}) - S(A_1 A_2 \cdots A_{2k +1}) \geq 0\,,
\end{equation}
where addition is computed mod $2k+1$. For $k=1$, \eqref{eq:cyclic} reduces to the MMI inequality \eqref{eq:mmi} which is the first member of this family.

Many mysteries surround these inequalities. The aim of this paper is to make progress towards two questions:

\begin{enumerate}
    \item What is their precise physical meaning? What are they telling us about gravity or the dual CFT?
    \item How general are they? Do they characterize a much larger class of physical systems than just those dual to Einstein gravity?
\end{enumerate}
 
Question (2) may seem to have, at first sight, a simple answer: since these inequalities are violated when $1/N$ quantum corrections are included, they should only be valid in the context of Einstein gravity. However, there is evidence that HEIs may hold for a larger class of theories. Examples include higher-derivative gravity theories, random tensor networks \cite{Hayden:2016cfa}, random states of large systems, ground states of topological theories \cite{Levin_2006, Kitaev:2005dm}, ground states and thermal states of sparse SYK models \cite{Iizuka:2024die}, and ground states of large-$c$ two-dimensional CFTs with vacuum block dominance \cite{Hartman:2013mia}. All of these examples share one thing in common: a large number of degrees of freedom. It has been conjectured therefore, that these inequalities may characterize typical states of large $N$ theories broadly defined. See \cite{Grimaldi:2026qhv} for a more comprehensive overview. 

In this paper we entertain this possibility. Motivated by the recurring theme that properties of entanglement teach us new aspects about quantum field theory, we ask the question: \emph{what do holographic entropy inequalities teach us about general large $N$ conformal field theories?} Answering this will bring us closer to resolving question (1) above. We will find sharp necessary conditions on the CFT, which we explain in more detail below. Armed with these constraints we can then test question (2) much more sharply: \emph{what kind of CFTs have these properties?} 

Constraining the space of large $N$ conformal field theories has a long history. In $d = 2$, the authors \cite{Hartman:2014oaa} find necessary and sufficient conditions for large-$c$ CFTs to have a thermal phase structure of 3d gravity. In particular, it was shown that if the CFT had a sparse light spectrum then the vacuum free energy dominates below the Hawking-Page phase transition between thermal AdS and the BTZ black hole \cite{ Witten:1998zw}. Always in $d=2$, a sharp statement about entanglement entropies of large-$c$ CFTs obeying this sparsity condition can be made \cite{Hartman:2013mia}. For such theories, the Renyi entropies for a number of disjoint intervals in the vacuum can be computed, and their $n\to 1$ limit equal to a sum of lengths of AdS$_3$ geodesics, agreeing with the RT formula. (See also the original paper \cite{Headrick:2010zt} where the case for two intervals was derived). Beyond theories dual to Einstein gravity, examples of theories of this type are symmetric-product orbifold theories. The assumption of sparsity of light operators translate to the fact that the conformal block expansion for the Renyi entropy is dominated by the vacuum block. For $d>2$, the requirements for large $N$ CFT to have the phase structure of Einstein gravity are stricter in that the entire spectrum is required to be sparse for theories defined on spatial tori \cite{Belin:2016yll}, and to our knowledge not much is known about entanglement entropy. 

The goal of our paper is similar to the above works but from the opposite angle. We do not assume anything about the spectrum of our CFT, nor we assume that its phase structure needs to match the leading Einstein gravity behavior. Instead, we assume that its entanglement structure is compatible with the dihedral family of holographic entropy inequalities \eqref{eq:cyclic} at large $N$. We find the following:
\begin{enumerate}
    \item In sec. \ref{sec:vacuum-dom}, under the above assumption, we prove that for any $d$-dimensional CFT in a thermal state, its free energy is dominated by its vacuum value whenever $\beta > \beta^{\star}$, where $\beta^{\star}$ is some threshold value which depends on the spacetime dimension.
    \item In sec. \ref{sec:conformal-bound}, we specialize to $d = 3$ and find that the above assumption implies that the ratio between the coefficient $C_T$ of the stress-tensor two-point function and the sphere's free energy $F_0 = -\log Z[\mathbb{S}^3]$ must satisfy the bound
    \begin{equation}
        \frac{C_T}{F_0} \leq \frac{3}{\pi^3}.
    \end{equation}
\end{enumerate}

Both computations are purely field theoretic with no assumption of a gravity dual.

We end the paper in sec. \ref{sec:applications} where we explore some of the consequences of these constraints. In particular, we test our two predictions for $3d$ ABJM theory and $\mathcal{N} = 4$ SYM in $d = 4$ in the planar (large $N$, weak coupling) limit, a regime where the bulk is very stringy and we do not have a notion of RT formula. We find that neither theory violates our necessary constraints. 

\section{Vacuum dominance from HEIs}\label{sec:vacuum-dom}

We begin by stating our assumptions. We consider $d$-dimensional CFT at large $N$ defined on the cylinder $\mathbb{R} \times \mathbb{S}^{d-1}$ with unit radius. Here $N$ should be understood as the effective inverse Planck constant of the theory, e.g. $c$ for $d = 2$ or $N^2$ for $d = 4$ SYM. Therefore we consider only the leading contributions in $N$ to entanglement entropy, and we assume that the dihedral family \eqref{eq:cyclic} is satisfied in this limit. 

Consider a thermal state on $\mathbb{S}^{d-1}$ at inverse temperature $\beta$. Divide the sphere equally into $2k + 1$ lunes with opening angles $\alpha_k = \frac{2\pi}{2k+1}$ and consider the dihedral inequality \eqref{eq:cyclic}, i.e.
\begin{equation}
    \sum_{i = 1}^{2k + 1} S(A_iA_{i+1}\dots A_{i+k}) - \sum_{i = 1}^{2k + 1} S(A_iA_{i+1}\dots A_{i+k-1}) - S(A_1 A_2 \cdots A_{2k +1}) \geq 0.
\end{equation}
The entropy of a lune depends only on its opening angle, so the above reduces to
\begin{equation}
(2k+1)\left(S((k+1)\alpha_k) - S(k\alpha_k)\right) - S_{\text{thermal}} \geq 0.
\end{equation}
Noting that $k\alpha_k = \pi - \alpha_k/2$, we have
\begin{equation}
(2k+1)\left(S(\pi + \alpha_k/2) - S(\pi - \alpha_k/2)\right) - S_{\text{thermal}} \geq 0.
\end{equation}
Taking the large $k$ limit, the above inequality becomes an inequality on the first derivative of the entropy of the hemisphere:
\begin{equation}\label{eq:lower-bound}
    S_{\text{thermal}} \leq 2\pi S'(\pi).
\end{equation}
We are now going to upper bound $S'(\pi)$. Consider the relative entropy of a lune of opening angle $\alpha$ between the thermal state $\rho_\beta$ and the vacuum $\sigma$ on the sphere
\begin{equation}
    S_{\alpha}(\rho_{\beta} |\sigma) = \Delta \langle K_{\alpha}\rangle - \Delta S_{\alpha},
\end{equation}
where $K_{\alpha} = -\log\sigma_{\alpha}$ is the modular Hamiltonian of the lune in the vacuum and $\Delta S_{\alpha} = S_{\beta}(\alpha) - S_{\sigma}(\alpha)$.
Consider two angles $\alpha \leq \alpha'$, monotoniciy of relative entropy under inclusion gives
\begin{equation}
    S_{\alpha}(\rho_{\beta} |\sigma) \leq S_{\alpha'}(\rho_{\beta} |\sigma)\,,
\end{equation}
when $\alpha'$ is a small deviation from $\alpha$ this gives us the differential inequality
\begin{equation}\label{eq:upper-bound}
    \frac{\partial}{\partial \alpha} \Delta S_{\alpha} \leq \frac{\partial}{\partial \alpha} \Delta \langle K_{\alpha} \rangle.
\end{equation}
Since $S_{\text{vacuum}}'(\pi) = 0$ by purity, we can combine \eqref{eq:lower-bound} with \eqref{eq:upper-bound}, obtaining
\begin{equation}\label{eq:lower-upper-bound}
    S_{\text{thermal}} \leq 2\pi S'(\pi) \leq 2\pi \frac{\partial}{\partial \alpha}  \Delta \langle K_{\alpha}\rangle\bigg|_{\alpha = \pi}
\end{equation}
The modular Hamiltonian for a \emph{spherical cap} region extending from the north-pole to the latitude $\theta_0$ was computed in \cite{Herzog:2014fra} to be
\begin{equation}
    K^{\text{cap}}(\theta_0) = 2\pi \int_{0}^{\theta_0}\int_{\mathbb{S}^{d-2}} T_{tt}(p) \frac{\cos\theta - \cos{\theta_0}}{\sin\theta_0}\sin^{d-2}\theta\,  \text{Vol}(\mathbb{S}^{d-2}) \, \text{d}\theta .
\end{equation}
We note that a spherical cap and a lune define the same shape, the hemisphere, when $\theta_0 = \pi/2$ and $\alpha = \pi$ respectively. We can compute the first variation of the above modular Hamiltonian for the hemisphere $\theta_0 = \pi/2$ under a cap-variation. We have
\begin{equation}
    \frac{\partial}{\partial \theta_0} K^{\text{cap}}(\theta_0)\bigg|_{\theta_0 = \pi/2}  = 2\pi \int_{0}^{\pi/2} \int_{\mathbb{S}^{d-2}} T_{tt}(p) \sin^{d-2}\theta\,  \text{Vol}(\mathbb{S}^{d-2}) \, \text{d}\theta
\end{equation}
This is an integral of the local energy density $T_{tt}$ over the entire hemisphere, therefore we have
\begin{equation}\label{eq:cap-var}
\frac{\partial}{\partial \theta}\Delta \langle K^{\text{cap}}\rangle\bigg|_{\theta = \pi/2} = \frac{\partial}{\partial \theta} \langle K^{\text{cap}}\rangle_{\beta}\bigg|_{\theta = \pi/2}  - \frac{\partial}{\partial \theta} \langle K^{\text{cap}}\rangle_{\text{vac}}\bigg|_{\theta = \pi/2} = \frac{2\pi}{2} \left(E_{\beta} - E_{\text{vac}}\right) := \pi \Delta E_{\beta}
\end{equation}
where $E_{\beta}$ and $E_{\text{vac}}$ are the thermal and vacuum energies of the whole sphere respectively.

\begin{figure}
    \centering
    \includegraphics[width=0.85\linewidth]{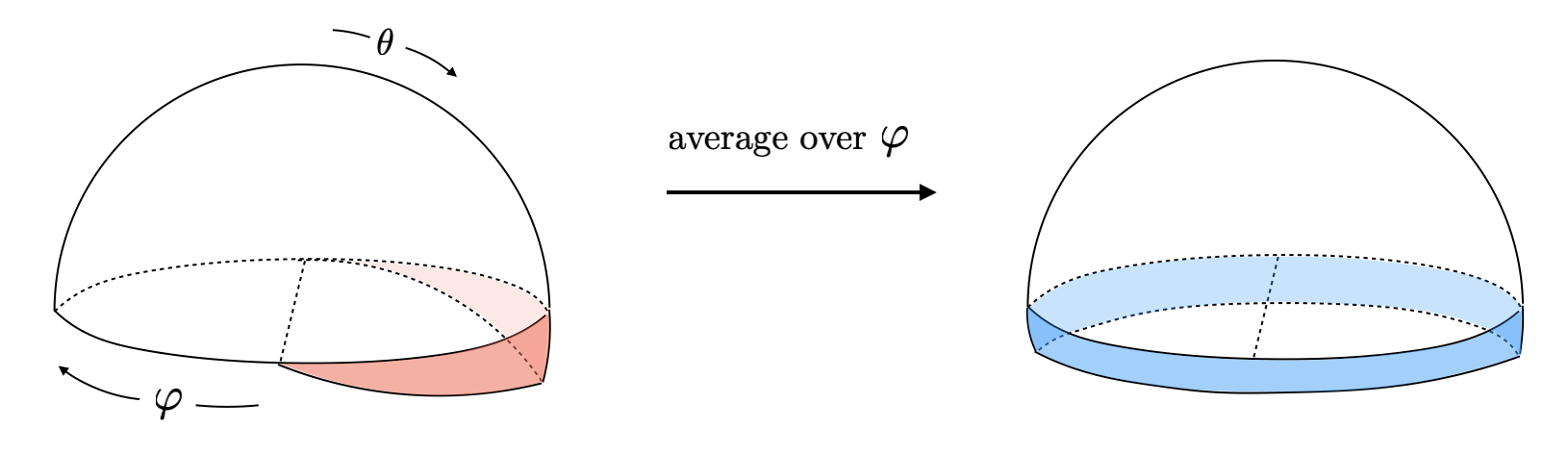}
    \caption{Example in $d = 3$. A single wedge deformation $\text{d}\alpha$ deforms the equator by $\sin\varphi \text{d}\alpha$ for $\varphi \in [0,\pi]$ and $0$ for $\varphi \in [\pi,2\pi]$. Rotating these deformations around the equatorial $S^1$ and averaging gives a cap deformation but with angle $\text{d}\theta_{\text{cap}} = \text{d}\alpha/\pi$.}
    \label{fig:deformation}
\end{figure}

The above result concerns the first order response of an hemisphere under a cap-deformation, whereas in \eqref{eq:lower-upper-bound} we are interested in a wedge-deformation. But we can relate the two by noting that since the first-order response $\delta K[f]$ is linear in the wedge-variation $f$, and since the hemisphere and state are rotationally invariant, we can average over rotated wedge-deformations $f_{i}$ with $i \in \Omega_{d-2}$, i.e.
\begin{equation}
    \delta K[f] = \frac{1}{\text{Vol}(\mathbb{S}^{d-2})}\int_{\mathbb{S}^{d-2}}\delta K[f_{i}] = \delta K\left[\frac{1}{\text{Vol}(\mathbb{S}^{d-2})}\int_{\mathbb{S}^{d-2}} f_{i}\right].
\end{equation}
where $f_{i}$ is a wedge-deformation by an angle $\text{d}\alpha$. Integrating over the entire $\mathbb{S}^{d-2}$ simply gives $\text{Vol}(S^{d-1})\frac{\text{d}\alpha}{2\pi}$, so this is equivalent to a single cap-deformation but with angle
\begin{equation}
    \text{d}\theta_{\text{cap}} = \frac{1}{2\pi}\frac{\text{Vol}(\mathbb{S}^{d-1})}{\text{Vol}(\mathbb{S}^{d-2})} \text{d}\alpha.
\end{equation}
See Figure \ref{fig:deformation} for a visual aid of the geometric argument. The first order variation on the modular Hamiltonian for a wedge-deformation is thus related to the one for a cap-deformation by
\begin{equation}\label{eq:wedge-var}
    \frac{1}{2\pi}\frac{\text{Vol}(\mathbb{S}^{d-1})}{\text{Vol}(\mathbb{S}^{d-2})}\frac{\partial}{\partial \theta}\Delta \langle K^{\text{cap}}\rangle\bigg|_{\theta = \pi/2} = \frac{\partial}{\partial \alpha}\Delta \langle K_{\alpha}\rangle\bigg|_{\alpha = \pi}.
\end{equation}
Then using \eqref{eq:lower-upper-bound}, \eqref{eq:cap-var}, and \eqref{eq:wedge-var} we have
\begin{equation}\label{eq:final-bound}
    S_{\text{thermal}} \leq 2\pi S'(\pi) \leq  \frac{\text{Vol}(\mathbb{S}^{d-1})}{\text{Vol}(\mathbb{S}^{d-2})}\frac{\partial}{\partial \theta}\Delta \langle K^{\text{cap}}\rangle\bigg|_{\theta = \pi/2} = \pi  \frac{\text{Vol}(\mathbb{S}^{d-1})}{\text{Vol}(\mathbb{S}^{d-2})} \Delta E_{\beta}.
\end{equation}
Decomposing the partition function into its vacuum and excited contribution, $\log{Z} = -\beta E_0 + \log{\hat{Z}}$ with $\hat{Z} = 1+ \sum_{n>0}e^{-\beta(E_n - E_0)}$, we have
\begin{equation}
S_{\text{thermal}} = (1  -\beta \partial_{\beta})\log{Z} = \log{\hat{Z}} + \beta \Delta E_\beta
\end{equation}
and so the bound \eqref{eq:final-bound} on the the thermal entropy becomes
\begin{equation}\label{eq:final-inequality}
    \log \hat{Z} + \left(\beta - \pi \frac{\text{Vol}(\mathbb{S}^{d-1})}{\text{Vol}(\mathbb{S}^{d-2})}\right)\Delta E_{\beta} \leq 0.
\end{equation}
The above inequality should be understood at leading order in $N$, and therefore, whenever
\begin{equation}
    \beta > \beta^{\star} = \pi \frac{\text{Vol}(\mathbb{S}^{d-1})}{\text{Vol}(\mathbb{S}^{d-2})}\,,
\end{equation}
the inequality \eqref{eq:final-inequality} is true only if $\log{\hat{Z}}$ and $\Delta E_{\beta}$ are subleading in $N$. In other words, the free energy of the theory is dominated by its vacuum value,
\begin{equation}
    F = - \log{Z} = \beta E_0 + \text{subleading in } N,
\end{equation}
whenever $\beta > \beta^{\star}$. For $d = 2,3,4,5,\infty$ we have $\beta^{\star} = \pi^2$, $2\pi$, $\pi^2/2$, $4\pi/3$, and $0$ respectively.

\section{A conformal bound on CFT$_3$s}\label{sec:conformal-bound}
\begin{figure}
    \centering
    \includegraphics[width=0.3\linewidth]{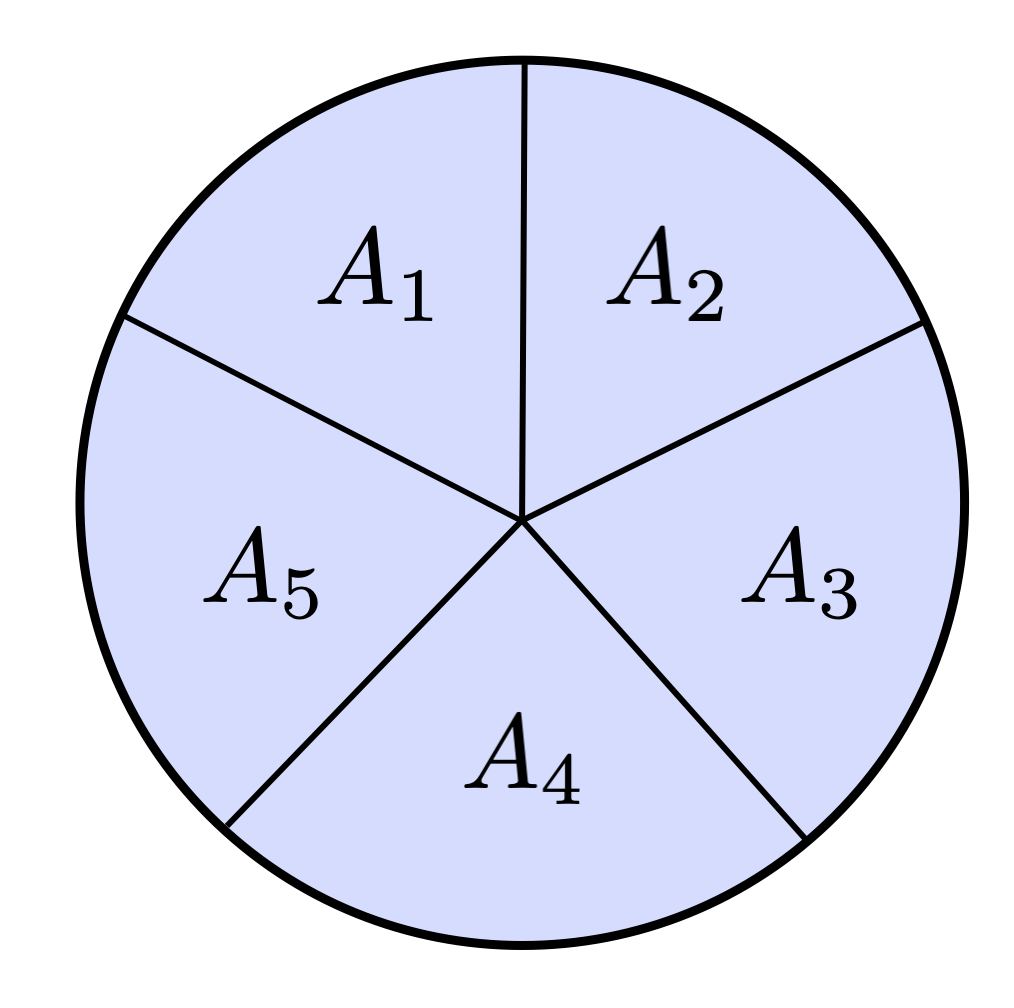}
    \caption{Configuration of regions considered in the proof. Here we show the case for $k = 2$, where the disk is split into 5 equal wedges with opening angle $\theta = 2\pi/5$.}
    \label{fig:disk}
\end{figure}

In this section we specialize to $d = 3$ and consider a CFT$_3$ in the vacuum on the plane. Once again, we assume that the the theory obeys the dihedral family \eqref{eq:cyclic} of holographic entropy inequalies at large $N$.\footnote{We note that the argument of this section goes through without assuming large $N$. We however interpret these results only at large $N$ since we know that the dihedral family of HEIs is generically violated away from large $N$.} The geometric setup for this part will be a disk with radius $R$, divided into $2k+1$ equal wedges each with opening angle $\theta_k = \frac{2\pi}{2k+1}$. We associate each wedge to each region $A_i$ in order, as shown in Figure \ref{fig:disk}.

In $d = 3$, the entanglement entropy for a disk of radius $R$ reads
\begin{equation}
    S(\text{disk}) = \alpha \frac{2\pi R}{\epsilon} - F_0\, ,
\end{equation}
where $\epsilon$ is a UV regulator, $\alpha$ is a non-universal coefficient and $F_0 = -\log Z[\mathbb{S}^3]$ the sphere's free energy. For a wedge with opening angle $\theta$, corner terms that scale logarithmically with the UV regulator must be included \cite{Casini:2006hu}, and its entanglement entropy reads
\begin{equation}\label{eq:ee-wedge}
    S(R, \theta) = \alpha \frac{R(2+\theta)}{\epsilon} - [a(\theta) + 2a(\pi/2)]\ln\frac{R}{\epsilon} + b(\theta)\,.
\end{equation}
The area-law term now contains the contribution from the two radii, the second term accounts for the contribution from the three corners ($a(\theta)$ contributes to the $\theta$-corner between the two radii, while $2a(\pi/2)$ to the two $\pi/2$ corners between each radius and the perimeter of the disk), and the last term is again a finite piece which depends on the opening angle only. The function $a(\theta)$ is a universal quantity, being related to $C_T$ \cite{Bueno:2015rda, Faulkner:2015csl}, and enjoys the symmetry $a(\theta) = a(2\pi - \theta)$. Specifically, the dependence of $a(\theta)$ with $C_T$ is explicit near $\theta \sim \pi$, where the function behaves as $a(\theta) \sim \sigma (\pi-\theta)^2$ with $\sigma = \pi^2 C_T/24$.

The procedure now is very similar to what done in sec. \ref{sec:vacuum-dom}. Computing the dihedral inequality on this configuration gives us
\begin{equation}
    (2k + 1) S(\pi + \theta_k/2) - (2k+1)S(\pi - \theta_k/2) - S(\text{disk}) \geq 0.
\end{equation}
The $\theta_k$-independent terms of $S$ cancel from the subtraction between $S(\pi + \theta_k/2)$ and $S(\pi - \theta_k/2)$, so we have
\begin{equation}
S(\pi + \theta_k/2) - S(\pi-\theta_k/2) = \alpha \frac{R \theta_k}{\epsilon} - [a(\pi + \theta_k/2) - a(\pi - \theta_k/2)]\ln\frac{R}{\epsilon} + b(\pi + \theta_k/2) - b(\pi - \theta_k/2)
\end{equation}
Using the fact that $a(\theta) = a(2\pi-\theta)$, we have $a(\pi + \theta_k/2) = a(2\pi - \pi - \theta_k/2) = a(\pi - \theta_k/2)$, so those terms cancel each other identically in the above expression, obtaining
\begin{equation}
S(\pi + \theta_k/2) - S(\pi-\theta_k/2) = \alpha \frac{R\theta_k}{\epsilon} + b(\pi + \theta_k/2) - b(\pi - \theta_k/2).
\end{equation}
Putting this back into the inequality, we have
\begin{equation}
    (2k + 1) \left[S(\pi + \theta_k/2) - S(\pi - \theta_k/2)\right] - S(\text{disk}) =  F_0 +(2k+1)\left(b(\pi + \theta_k/2) - b(\pi - \theta_k/2)\right) \geq 0
\end{equation}
Taking the large $k$ limit, the above reduces to
\begin{equation}\label{eq:b-bound}
     F_0 + 2\pi b'(\pi) \geq 0\,.
\end{equation}
To compute $b'(\pi)$, we look at the first order change in entropy under an angle deformation in two different conformal frames.

\paragraph{Half-disk} Set $R = 1$. Differentiating \eqref{eq:ee-wedge} with respect to an angle deformation of the wedge, and evaluating it at the half-disk $\theta = \pi$, we find
\begin{equation}\label{eq:half-disk-var}
S'(\text{half-disk}) = \frac{\alpha}{\epsilon} + b'(\pi).
\end{equation}
where we have made use of the fact that $a(\theta) \sim (\theta-\pi)^2$ when $\theta \sim \pi$, which gives us $a'(\pi) = 0$.

\paragraph{Quadrant}

\begin{figure}
    \centering
    \includegraphics[width=0.8\linewidth]{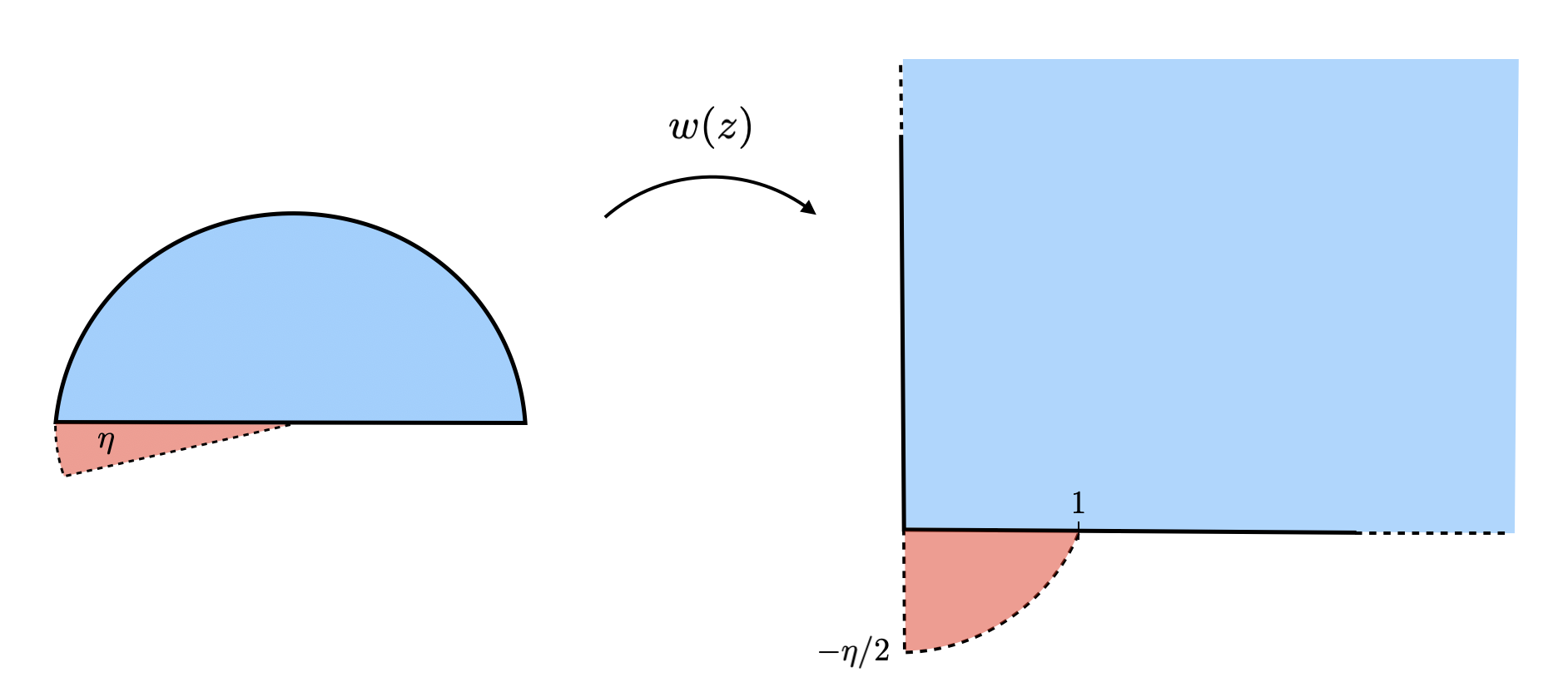}
    \caption{Conformal map of the half-disk (blue) to the infinite quadrant with its angular deformation (red).}
    \label{fig:conformal-map}
\end{figure}

Consider the conformal transformation
\begin{equation}\label{eq:conformal-map}
    w(z) = \frac{1+z}{1-z} \,,
\end{equation}
which maps the upper half-disk to the first quadrant. Specifically, the diameter of the disk is mapped to the real axis, with the points $z = -1$, $z = 0$ and $z = 1$ to the points $w = 0$, $w = 1$ and $w = +\infty$ respectively, and the semi-circle to the imaginary axis, with the points $\theta = 0$ and $\theta = \pi$ to the points $w = 0$ and $w = i\infty$ respectively. 

The small angular deformation that sends the half-disk radial points $r \in [-1,0]$ to $re^{i\eta}$, under the above conformal map, corresponds to the deforming the $x$-axis of the quadrant downward by $-\tan\frac{\eta}{2} + \tan\frac{\eta}{2}x^2$ with support on $0<x<1$. See Figure \ref{fig:conformal-map}.

To compute the first order change in the entanglement entropy under the deformation, we make use of the first variation formula 
\begin{equation}\label{eq:first-order}
    \delta S = \int_{\partial A} \text{d}s \,  \varphi(s) D(s) 
\end{equation}
where $\varphi(s)$ is a scalar field on $\partial A$, implementing the normal deformation, and $D$ is the 1-pt function of the so-called displacement operator. For the purposes of our computation, we don't need to know the details of how the above formula comes about; we will sketch briefly its origins and defer the interested readers to the original paper \cite{Bianchi:2015liz}. The main idea comes from the twist operator perspective of the replica trick computation of entanglement entropy: instead of computing the path integral on an $n$-sheeted cover obtained from gluing the $n$ copies along $\partial A$ we can instead compute the path integral of $n$ CFTs defined on a single copy with an extended $1$-dimensional twist operator localized on $\partial A$. Deforming the entangling surface corresponds to transforming the operator under a diffeomorphism, which is implemented by the displacement operator.

As mentioned, the details on how to properly define $D$ do not matter for our setup since scale invariance of the quadrant enforces $D(r) = f(\theta)/r^2$, and we can fix the precise form of $f(\theta)$ by matching the result obtained using \eqref{eq:first-order} to the known angular deformation of an infinite wedge with opening $\theta$:
\begin{equation}
    S'(\theta) = -a'(\theta)\ln\frac{L}{\epsilon}\,.
\end{equation}
Doing so, one finds $f(\theta) = a'(\theta)$. 

Now we have all the ingredients to compute the first order variation on the quadrant. There will be two contributions: one from the added vertical displacement along the $y$ axis, which is computed simply by integrating the segment with the correctly transformed UV-cuttoff $\tilde{\epsilon} = |w'(z)| \epsilon$, i.e.
\begin{equation}
    \delta S_{y} =  \alpha\int \frac{\text{d}w}{\tilde{\epsilon}} = \frac{1}{\epsilon}\int_{-\tan(\eta/2)}^{0} \frac{2\,\text{d}y}{1+y^2} = \alpha\frac{\eta}{\epsilon}
\end{equation}
which is the UV divergent term that matches the one in \eqref{eq:half-disk-var}, and one from a contribution $\delta S_{\text{arc}}$  due to the deformation\footnote{We are working with the relative normal deformation with respect to the vertical shift by $-\tan\frac{\eta}{2}$.} $\tan\frac{\eta}{2}x^2 $ for $0<x<1$ and $\tan\frac{\eta}{2}$ for $x>1$. For this, which is a normal deformation, we use \eqref{eq:first-order} to obtain
\begin{equation}
    \delta S_{\text{arc}} = \tan\frac{\eta}{2}a'\left(\frac{\pi}{2}\right)\left[\int_{0}^{1} \text{d}x + \int_{1}^{\infty} \frac{\text{d}x}{x^2} \right]=  \eta a'(\pi/2) + O(\eta^3)\,, 
\end{equation}
which is a finite contribution that can be directly match to the half-disk answer \eqref{eq:half-disk-var}, obtaining $b'(\pi) = a'(\pi/2)$.

The first derivative $a'(\theta)$ of the corner term is only directly related to $C_T$ in the $\theta \to \pi$ limit, however luckily for us, it is in general upper-bounded by it \cite{Bueno:2015ofa}:
\begin{equation}
    a'(\theta) \leq - \frac{\pi^2 C_T}{6}\cot\frac{\theta}{2}.
\end{equation}
Setting $\theta = \pi/2$ in the above, and combining it with \eqref{eq:b-bound} we obtain our desired bound
\begin{equation}\label{eq:bound}
    \frac{C_T}{F_0} \leq \frac{3}{\pi^3}.
\end{equation}
We note that our bound fits with what proposed by the authors of \cite{Bueno:2023gey}, who conjectured a universal bound on $C_T/F_0$ for any 3-dimensional conformal field theory:
\begin{equation}\label{eq:cft3-bound}
    \frac{C_T}{F_0} \leq \frac{3}{4\pi^2 \log 2 - 6\zeta[3]} \approx 0.14887.
\end{equation}

\section{Applications and conclusions}\label{sec:applications}

In this paper we derived two constraints on CFTs solely from the validity of the dihedral family \eqref{eq:cyclic} of holographic entropy inequalities at large $N$.

\begin{enumerate}
    \item A constraint on the free energy of the theory: any $d$-dimensional CFT obeying the dihedral family in a thermal state at large $N$ must have a free energy dominated by its vacuum value for any temperature $T$ below some threshold, i.e.
\begin{equation}
    F = \beta E_0 + \text{subleading in }N, \quad \forall\, T < \frac{1}{\pi}\frac{\text{Vol}(\mathbb{S}^{d-2})}{\text{Vol}(\mathbb{S}^{d-1})}\,.
\end{equation}
\item A bound on the ratio of two universal data of the theory: $d = 3$ CFTs obeying the dihedral family \eqref{eq:cyclic} in the vacuum must satisfy
\begin{equation}
    \frac{C_T}{F_0} \leq \frac{3}{\pi^3}.
\end{equation}
\end{enumerate}

What should be the interpretation of these constraints? Restricting to theories dual to Einstein holography, we know they exhibit vacuum dominance below the Hawking-Page temperature $T_{\text{HP}} = (d-1)/2\pi$ which is stricter than our bound and (ii) we know the ratio $C_T/F_0$ exactly there
\begin{equation}\label{eq:einstein-bound}
    \frac{C_T}{F_0}\bigg|_{\text{Einstein-holo}} = \frac{6}{\pi^4} \approx 0.06159\,.
\end{equation}
However, vacuum dominance in large $N$ CFTs has been derived under the assumption of sparsity of the spectrum of the theory \cite{Hartman:2014oaa, Belin:2016yll}. Since we make no such assumption, our result can be understood as a purely entropic justification for it coming from the entanglement structure of the theory.

Away from theories dual to Einstein holography, these constraints instead give us necessary (but not sufficient) conditions on the space of CFTs to have an entanglement structure compatible with that of standard holographic states. (We stress the use of the word "compatible" rather than "identical" because the bound was derived only from the dihedral family of HEIs and not from the entire holographic entropy cone.) In particular, they allow us to sharply attack the question of whether the HEIs characterize a much larger class of many-body quantum systems (holographic, or not) than just those dual to Einstein holography. 

A good testing ground then is to take known top-down models in the planar limit, i.e. at large $N$ but at weak 't Hooft coupling. In this regime the bulk is very stringy, so the notion of an RT-like formula is completely lost. These constraints allow us to probe a corner of parameter space that would otherwise be completely inaccessible.

\subsection{$3d$ ABJM theory}
Here we consider a top-down example: $\mathcal{N} = 6$ ABJM theory with gauge group $U(N)_{k}\times U(N)_{-k}$ which is holographically dual to M-theory on AdS$_4\times S^7/\mathbb{Z}_k$, with $\pm k$ being the Chern-Simons levels. In the $N \to \infty$ limit, for any $k$, the bulk theory reduces to two-derivative supergravity, which includes Einstein's holography as its purely gravitational sector. Here we are interested to study the planar limit of the theory, which is achieved by taking both $N,k \to \infty$ while keeping the 't Hooft coupling
\begin{equation}
    \lambda = \frac{N}{k}
\end{equation}
fixed. For $\lambda \gg 1$ the bulk is semiclassical and the action contains higher-curvature corrections to Einstein gravity. In that case the ratio $C_T/F_0$ deviates from the Einstein value \eqref{eq:einstein-bound} by $1/\lambda$ corrections and therefore it will not violate our bound \eqref{eq:bound}.

We are therefore interested in considering the more extreme limit of weakly coupled planar ABJM, which correspond to taking $\lambda \ll 1$ and working in perturbation theory around it. In this limit the bulk is very stringy, while the large $N$ limit suppresses any quantum effects.

The values of $C_T$ and $F_0$ for this limit were computed in \cite{Gorini:2020new} and \cite{Drukker:2010nc} respectively, and are
\begin{align}
    C_T &= N^2\left(\frac{3}{2 \pi^2} - \frac{1}{2} \lambda^2 + O(\lambda^4) \right)\,,\\
    F_0 &=  N^2 \left( \log\left(\frac{2}{\pi \lambda}\right) + \frac{3}{2}  + \frac{\pi^2}{9}\lambda^2 + O(\lambda^4) \right).
\end{align}
Putting everything together we find that the ratio never violates the bound for small $\lambda$, with $C_T/F_0 \to 0$ as $\lambda \to 0$.

For the constraint on the vacuum free energy, the paper \cite{Ekhammar:2023cuj}
shows that $3d$ weakly coupled ABJM undergoes a Hagedorn phase transition at a temperature $T_H$ from a cold phase where the free energy is $O(1)$ to a hot phase where the free energy is order $O(N^2)$. Therefore, it is the perfect testing ground for our constraint. In $d = 3$ our critical temperature is
\begin{equation}
    T^{\star}_{3d} = \frac{1}{\pi}\frac{\text{Vol}(\mathbb{S}^1)}{\text{Vol}(\mathbb{S}^2)} = \frac{1}{2\pi} \approx 0.16,
\end{equation}
whereas the free energy of ABJM theory starts being vacuum-dominated below $T_H = 0.28$. 

All in all, we conclude that weakly coupled planar ABJM does not come close to violating either constraint, and it therefore is compatible with an RT-entanglement structure. 

\subsection{$4d$ $\mathcal{N} = 4$ SYM}

We can do a similar analysis for planar $4d$ $\mathcal{N} = 4$ SYM at weak coupling. The Hagedorn phase transition for weakly coupled $\mathcal{N} = 4$ SYM in $d = 4$ on an $\mathbb{S}^3$ of unit radius was computed in \cite{Aharony:2003sx}. Below the Hagedorn temperature the free energy is $O(1)$ and we have a gas of "glueballs", whereas above it we have a plasma with energy $O(N^2)$. For $d = 4$ we have 
\begin{equation}
    T^{\star}_{4d} = \frac{1}{\pi}\frac{\text{Vol}(\mathbb{S}^2)}{\text{Vol}(\mathbb{S}^3)} = \frac{2}{\pi^2} \approx 0.20,
\end{equation}
whereas free $\mathcal{N} =4$ SYM has $T_H = 0.38$. Once again, our bound is satisfied. 

\subsection{$3d\,\, O(N)$ Wilson-Fisher fixed point}
The authors of \cite{Bueno:2023gey} (and refs therein) provide the ratio $C_T/F_0$ for the large $N$ limit of the Wilson-Fisher fixed
point of the $O(N)$ model. The leading contribution is that of the free scalar, which saturates the conjectured bound \eqref{eq:cft3-bound}, and so violates our bound \eqref{eq:bound}. This result agrees with what found by the authors of \cite{Whitsitt:2016irx}, who computed the leading contribution to the entanglement entropy for this theory at large $N$ observing that it is given by that of $N$ free scalar fields which are known to violate the HEIs. At large $N$, this theory is dual to minimal Vasiliev higher-spin gauge theory in AdS$_4$ \cite{Klebanov:2002ja}.

\subsection{Phase transition in mutual information}
There is a striking consequence of the result from sec. \ref{sec:vacuum-dom}. Consider a $d$-dimensional CFT on $\mathbb{R}^{d}$ and two disjoint balls on the same time-slice. For $d = 2$, these will be two disjoint intervals $A$ and $B$, and in the seminal paper \cite{Headrick:2010zt} it was shown that in large-$c$ CFTs the mutual information $I(A:B) := S(A) + S(B) - S(AB)$ exhibits a phase transition as a function of their separation. This is beautifully realized holographically by the RT formula as the exchange of dominance between a connected phase (positive mutual information) and a disconnected phase (zero mutual information) of the RT surface for $S(AB)$. In $d>2$ the story is similar, but the phase transition is understood mostly from the gravitational side.

Recently, it was shown that in any $d$-dimensional CFT the mutual information for the two balls is bounded by the thermal partition function on $\mathbb{S}^{d-1}$, with an effective temperature which depends on the balls' cross ratio $x$ \cite{Jia:2026ujf}. (See also \cite{Ceyhan:2025qrj} for related previous work.) More precisely, the author of \cite{Jia:2026ujf} finds, using our notation from sec. \ref{sec:vacuum-dom}:
\begin{equation}\label{eq:jia-bound}
    I(A:B) \leq \log \hat{Z}(\beta_{\text{eff}}(x))\,,
\end{equation}
where $\beta_{\text{eff}}(x) = 4\,\text{arccosh}(x^{-1/2})$ is the effective inverse temperature. When the balls are infinitely far apart ($x\to 0$) $\beta_{\text{eff}}$ approaches infinity, whereas it approaches zero when the balls touch ($x\to1$). 

From our thermal threshold we have that the right-hand side of \eqref{eq:jia-bound} vanishes (to leading order in $N$) for all effective temperatures $\beta_{\text{eff}} > \beta^\star$. Since $I(A:B)\geq 0$ by subadditivity, our thermal threshold implies that for any $d$-dimensional CFT the mutual information of two balls vanishes at leading order in $N$ for 
\begin{equation}\label{eq:phase-transition}
  0 < x < \sech^2\frac{\beta^\star}{4}  
\end{equation}
i.e. for separations larger than some threshold value which is set by the thermal threshold for vacuum dominance.

This provides a field-theory description of the phase transition in mutual information for arbitrary dimensions, with no assumptions other than validity of the dihedral family of holographic entropy inequalities. Clearly since \eqref{eq:jia-bound} only bounds the mutual information, the threshold \eqref{eq:phase-transition} we found is only a conservative threshold for the connected-disconnected phase transition of entropy; the actual phase transition will have happened at separations smaller than that. We find however remarkable that one can place a bound on the critical separation for the phase transition in mutual information for any $d-$dimensional CFT at large $N$ by only imposing such a seemingly simple constraint.

We end by remarking that our result agrees with the large-$d$ exploration done in \cite{Colin-Ellerin:2019vst}, where the authors studied the phase transition of holographic mutual information in the large-$d$ limit of general relativity. They found that holographic correlations decouple in this limit. Indeed, for $d\to\infty$ we have
\begin{equation}
    \beta^{\star} = \pi \frac{\text{Vol}(\mathbb{S}^{d-1})}{\text{Vol}(\mathbb{S}^{d-2})} \to 0, \quad \text{sech}^2 \beta^{\star}/4 \to 1
\end{equation}
Hence, the mutual information vanishes for any separation of the balls. This is not only a check of agreement between our prediction and the results of \cite{Colin-Ellerin:2019vst}, but it is a sharp prediction for any large $N$ large $d$ CFT (possibly not dual to Einstein gravity) which obeys the dihedral inequalities. 

\subsection{Outlook}

We take these results as encouraging evidence in favor of the conjecture that holographic entropy inequalities characterize large $N$ field theories, at least those that admit an Einstein sector description in their strong coupling limit. However, one should be cautious since these constraints are only necessary. Further, the derivation of these constraints relied only on the dihedral family of HEIs, which is but a minuscule fraction of the whole set of constraints. 

We also take these results as suggesting that an holographic entanglement structure for large $N$ conformal field theories places drastic constraints on the spectrum of the theory. This is because our assumptions (large $N$, validity of dihedral inequalities) result in similar, albeit weaker, consequences (vacuum-dominated free energy at cold temperatures) as those found by \cite{Hartman:2014oaa} in $d = 2$, and by \cite{Belin:2016yll} in $d >2$, both of which required sparsity of the spectrum in addition to large $N$. It seems therefore plausible to conjecture that the holographic entropy cone knows intimately about the spectrum of large $N$ conformal field theories.

We view these results as a proof of concept for a much more ambitious program of understanding the implications that holographic entropy inequalities have on large $N$ CFTs. We are hopeful that the full power of the holographic entropy cone may teach us very interesting lessons about quantum field theory and quantum gravity. Even without full knowledge of the set of HEIs, we know that they are incredibly constrained by a series of mysterious combinatorial properties \cite{Grimaldi:2025jad, Grimaldi:2026lbq}. Could one extract information about the spectrum of the CFT from these combinatorial constraints?

Another interesting future direction is to use the two other known families of HEIs, the projective-plane and the toric inequalities \cite{Czech:2024rco}, to extract new constraints on large $N$ CFTs. One could perhaps explore different geometric constructions for the boundary regions that suit the symmetries of each family, or even consider large $N$ CFTs on spatial manifolds with different topologies.

Finally, it would be interesting to test these necessary constraints for a larger family of examples in more detail, such as bottom-up higher derivative theories and other conformal field theories at large $N$ (with, or without an Einstein gravity low-energy dual). A violation of our constraints at large $N$ would help understand whether these inequalities characterize large $N$ CFTs with an Einstein holographic dual at low-energies, or something else entirely.

\section*{Acknowledgments}
The author thanks Matt Headrick, Martin Sasieta, and Nico Valdes-Meller for useful comments and discussions. The author particularly wants to thank Matt Headrick for great insights that led to the exploration of sec. \ref{sec:vacuum-dom}. The author would like to thank Rob Myers and the Perimeter Institute for Theoretical Physics for hospitality and financial support during the writing of this work. 

\paragraph{Statement of AI use} The scientific vision and motivations that led to the project are the author's own. The proofs were carried out with the use of GPT 6.2 (Open AI), which were subsequently checked, independently carried out in detail, and modified by the author according to personal taste. No AI-generated text appears in this paper. GPT 6.2 was also useful for literature review. The author takes full responsibility for the accuracy of this work.

\bibliographystyle{jhep}
\bibliography{references}

@article{Faulkner:2015csl,
    author = "Faulkner, Thomas and Leigh, Robert G. and Parrikar, Onkar",
    title = "{Shape Dependence of Entanglement Entropy in Conformal Field Theories}",
    eprint = "1511.05179",
    archivePrefix = "arXiv",
    primaryClass = "hep-th",
    doi = "10.1007/JHEP04(2016)088",
    journal = "JHEP",
    volume = "04",
    pages = "088",
    year = "2016"
}

@article{Casini:2006hu,
    author = "Casini, H. and Huerta, M.",
    title = "{Universal terms for the entanglement entropy in 2+1 dimensions}",
    eprint = "hep-th/0606256",
    archivePrefix = "arXiv",
    doi = "10.1016/j.nuclphysb.2006.12.012",
    journal = "Nucl. Phys. B",
    volume = "764",
    pages = "183--201",
    year = "2007"
}

@article{Klebanov:2002ja,
    author = "Klebanov, I. R. and Polyakov, A. M.",
    title = "{AdS dual of the critical O(N) vector model}",
    eprint = "hep-th/0210114",
    archivePrefix = "arXiv",
    reportNumber = "PUPT-2053",
    doi = "10.1016/S0370-2693(02)02980-5",
    journal = "Phys. Lett. B",
    volume = "550",
    pages = "213--219",
    year = "2002"
}

@article{Whitsitt:2016irx,
    author = "Whitsitt, Seth and Witczak-Krempa, William and Sachdev, Subir",
    title = "{Entanglement entropy of the large $N$ Wilson-Fisher conformal field theory}",
    eprint = "1610.06568",
    archivePrefix = "arXiv",
    primaryClass = "cond-mat.str-el",
    doi = "10.1103/PhysRevB.95.045148",
    journal = "Phys. Rev. B",
    volume = "95",
    number = "4",
    pages = "045148",
    year = "2017"
}

@article{Colin-Ellerin:2019vst,
    author = "Colin-Ellerin, Sean and Hubeny, Veronika E. and Niehoff, Benjamin E. and Sorce, Jonathan",
    title = "{Large-$d$ phase transitions in holographic mutual information}",
    eprint = "1911.06339",
    archivePrefix = "arXiv",
    primaryClass = "hep-th",
    doi = "10.1007/JHEP04(2020)173",
    journal = "JHEP",
    volume = "04",
    pages = "173",
    year = "2020"
}

@article{Ceyhan:2025qrj,
    author = "Ceyhan, Fikret and Faulkner, Thomas",
    title = "{Bounds on CFT correlations from the thermal partition function}",
    eprint = "2510.24042",
    archivePrefix = "arXiv",
    primaryClass = "hep-th",
    month = "10",
    year = "2025"
}

@article{Bueno:2015ofa,
    author = "Bueno, Pablo and Witczak-Krempa, William",
    title = "{Bounds on corner entanglement in quantum critical states}",
    eprint = "1511.04077",
    archivePrefix = "arXiv",
    primaryClass = "cond-mat.str-el",
    doi = "10.1103/PhysRevB.93.045131",
    journal = "Phys. Rev. B",
    volume = "93",
    pages = "045131",
    year = "2016"
}

@article{Jia:2026ujf,
    author = "Jia, Hewei Frederic",
    title = "{Vacuum correlation measures as CFT partition functions}",
    eprint = "2611.34922",
    archivePrefix = "arXiv",
    primaryClass = "hep-th",
    month = "11",
    year = "2026"
}

@article{Headrick:2010zt,
    author = "Headrick, Matthew",
    title = "{Entanglement Renyi entropies in holographic theories}",
    eprint = "1006.0047",
    archivePrefix = "arXiv",
    primaryClass = "hep-th",
    reportNumber = "BRX-TH-619, BRX-TH 619",
    doi = "10.1103/PhysRevD.82.126010",
    journal = "Phys. Rev. D",
    volume = "82",
    pages = "126010",
    year = "2010"
}

@article{Grimaldi:2026lbq,
    author = "Grimaldi, Guglielmo and Headrick, Matthew and Hubeny, Veronika E. and Shteyner, Pavel",
    title = "{Combinatorial properties of holographic entropy inequalities}",
    eprint = "2601.09987",
    archivePrefix = "arXiv",
    primaryClass = "hep-th",
    month = "1",
    year = "2026"
}

@article{Ekhammar:2023cuj,
    author = "Ekhammar, Simon and Minahan, Joseph A. and Thull, Charles",
    title = "{The ABJM Hagedorn Temperature from Integrability}",
    eprint = "2307.02350",
    archivePrefix = "arXiv",
    primaryClass = "hep-th",
    reportNumber = "UUITP-16/23, MIT-CTP/5574",
    doi = "10.1007/JHEP10(2023)066",
    journal = "JHEP",
    volume = "10",
    pages = "066",
    year = "2023"
}

@article{Belin:2016yll,
    author = "Belin, Alexandre and de Boer, Jan and Kruthoff, Jorrit and Michel, Ben and Shaghoulian, Edgar and Shyani, Milind",
    title = "{Universality of sparse $d > 2$ conformal field theory at large $N$}",
    eprint = "1610.06186",
    archivePrefix = "arXiv",
    primaryClass = "hep-th",
    doi = "10.1007/JHEP03(2017)067",
    journal = "JHEP",
    volume = "03",
    pages = "067",
    year = "2017"
}

@article{Aharony:2003sx,
    author = "Aharony, Ofer and Marsano, Joseph and Minwalla, Shiraz and Papadodimas, Kyriakos and Van Raamsdonk, Mark",
    editor = "Doebner, H. D. and Dobrev, V. K.",
    title = "{The Hagedorn - deconfinement phase transition in weakly coupled large N gauge theories}",
    eprint = "hep-th/0310285",
    archivePrefix = "arXiv",
    reportNumber = "WIS-29-03-DPP",
    doi = "10.4310/ATMP.2004.v8.n4.a1",
    journal = "Adv. Theor. Math. Phys.",
    volume = "8",
    pages = "603--696",
    year = "2004"
}

@article{Hayden:2016cfa,
    author = "Hayden, Patrick and Nezami, Sepehr and Qi, Xiao-Liang and Thomas, Nathaniel and Walter, Michael and Yang, Zhao",
    title = "{Holographic duality from random tensor networks}",
    eprint = "1601.01694",
    archivePrefix = "arXiv",
    primaryClass = "hep-th",
    doi = "10.1007/JHEP11(2016)009",
    journal = "JHEP",
    volume = "11",
    pages = "009",
    year = "2016"
}

@article{Witten:1998zw,
    author = "Witten, Edward",
    editor = "Bergstrom, L. and Lindstrom, U.",
    title = "{Anti-de Sitter space, thermal phase transition, and confinement in gauge theories}",
    eprint = "hep-th/9803131",
    archivePrefix = "arXiv",
    reportNumber = "IASSNS-HEP-98-21",
    doi = "10.4310/ATMP.1998.v2.n3.a3",
    journal = "Adv. Theor. Math. Phys.",
    volume = "2",
    pages = "505--532",
    year = "1998"
}

@article{Hartman:2014oaa,
    author = "Hartman, Thomas and Keller, Christoph A. and Stoica, Bogdan",
    title = "{Universal Spectrum of 2d Conformal Field Theory in the Large c Limit}",
    eprint = "1405.5137",
    archivePrefix = "arXiv",
    primaryClass = "hep-th",
    reportNumber = "CALT-68-2889, RUNHETC-2014-07",
    doi = "10.1007/JHEP09(2014)118",
    journal = "JHEP",
    volume = "09",
    pages = "118",
    year = "2014"
}

@article{Herzog:2014fra,
    author = "Herzog, Christopher P.",
    title = "{Universal Thermal Corrections to Entanglement Entropy for Conformal Field Theories on Spheres}",
    eprint = "1407.1358",
    archivePrefix = "arXiv",
    primaryClass = "hep-th",
    doi = "10.1007/JHEP10(2014)028",
    journal = "JHEP",
    volume = "10",
    pages = "028",
    year = "2014"
}

@article{Kudler-Flam:2026abf,
    author = "Kudler-Flam, Jonah",
    title = "{Proof of the entropic $c$-theorem}",
    eprint = "2610.40874",
    archivePrefix = "arXiv",
    primaryClass = "hep-th",
    month = "10",
    year = "2026"
}

@article{Casini:2017vbe,
    author = "Casini, Horacio and Test{\'e}, Eduardo and Torroba, Gonzalo",
    title = "{Markov Property of the Conformal Field Theory Vacuum and the a Theorem}",
    eprint = "1704.01870",
    archivePrefix = "arXiv",
    primaryClass = "hep-th",
    doi = "10.1103/PhysRevLett.118.261602",
    journal = "Phys. Rev. Lett.",
    volume = "118",
    number = "26",
    pages = "261602",
    year = "2017"
}

@article{Zamolodchikov:1986gt,
    author = "Zamolodchikov, A. B.",
    title = "{Irreversibility of the Flux of the Renormalization Group in a 2D Field Theory}",
    journal = "JETP Lett.",
    volume = "43",
    pages = "730--732",
    year = "1986"
}

@article{Casini:2004bw,
    author = "Casini, H. and Huerta, M.",
    title = "{A Finite entanglement entropy and the c-theorem}",
    eprint = "hep-th/0405111",
    archivePrefix = "arXiv",
    doi = "10.1016/j.physletb.2004.08.072",
    journal = "Phys. Lett. B",
    volume = "600",
    pages = "142--150",
    year = "2004"
}

@article{Casini:2006es,
    author = "Casini, H. and Huerta, M.",
    editor = "Sola, Joan",
    title = "{A c-theorem for the entanglement entropy}",
    eprint = "cond-mat/0610375",
    archivePrefix = "arXiv",
    doi = "10.1088/1751-8113/40/25/S57",
    journal = "J. Phys. A",
    volume = "40",
    pages = "7031--7036",
    year = "2007"
}

@article{Casini:2012ei,
    author = "Casini, H. and Huerta, Marina",
    title = "{On the RG running of the entanglement entropy of a circle}",
    eprint = "1202.5650",
    archivePrefix = "arXiv",
    primaryClass = "hep-th",
    doi = "10.1103/PhysRevD.85.125016",
    journal = "Phys. Rev. D",
    volume = "85",
    pages = "125016",
    year = "2012"
}

@article{Bueno:2015rda,
    author = "Bueno, Pablo and Myers, Robert C. and Witczak-Krempa, William",
    title = "{Universality of corner entanglement in conformal field theories}",
    eprint = "1505.04804",
    archivePrefix = "arXiv",
    primaryClass = "hep-th",
    doi = "10.1103/PhysRevLett.115.021602",
    journal = "Phys. Rev. Lett.",
    volume = "115",
    pages = "021602",
    year = "2015"
}

@inproceedings{Grimaldi:2026qhv,
    author = "Grimaldi, Guglielmo and Headrick, Matthew",
    title = "{Holographic entropy inequalities {\&} typical states}",
    booktitle = "{Strings 2026}",
    eprint = "2610.33905",
    archivePrefix = "arXiv",
    primaryClass = "hep-th",
    reportNumber = "BRX-TH-6733",
    month = "10",
    year = "2026"
}

@article{Drukker:2010nc,
    author = "Drukker, Nadav and Marino, Marcos and Putrov, Pavel",
    title = "{From weak to strong coupling in ABJM theory}",
    eprint = "1007.3837",
    archivePrefix = "arXiv",
    primaryClass = "hep-th",
    reportNumber = "HU-EP-10-39",
    doi = "10.1007/s00220-011-1253-6",
    journal = "Commun. Math. Phys.",
    volume = "306",
    pages = "511--563",
    year = "2011"
}

@article{Gorini:2020new,
    author = "Gorini, Nicola and Griguolo, Luca and Guerrini, Luigi and Penati, Silvia and Seminara, Domenico and Soresina, Paolo",
    title = "{The topological line of ABJ(M) theory}",
    eprint = "2012.11613",
    archivePrefix = "arXiv",
    primaryClass = "hep-th",
    doi = "10.1007/JHEP06(2021)091",
    journal = "JHEP",
    volume = "06",
    pages = "091",
    year = "2021"
}

@article{Bianchi:2015liz,
    author = "Bianchi, Lorenzo and Meineri, Marco and Myers, Robert C. and Smolkin, Michael",
    title = "{R{\'e}nyi entropy and conformal defects}",
    eprint = "1511.06713",
    archivePrefix = "arXiv",
    primaryClass = "hep-th",
    reportNumber = "DESY-15-229",
    doi = "10.1007/JHEP07(2016)076",
    journal = "JHEP",
    volume = "07",
    pages = "076",
    year = "2016"
}

@article{Bueno:2023gey,
    author = "Bueno, Pablo and Casini, Horacio and Andino, Oscar Lasso and Moreno, Javier",
    title = "{Conformal Bounds in Three Dimensions from Entanglement Entropy}",
    eprint = "2307.05164",
    archivePrefix = "arXiv",
    primaryClass = "hep-th",
    doi = "10.1103/PhysRevLett.131.171601",
    journal = "Phys. Rev. Lett.",
    volume = "131",
    number = "17",
    pages = "171601",
    year = "2023"
}

@article{Grimaldi:2025jad,
    author = "Grimaldi, Guglielmo and Headrick, Matthew and Hubeny, Veronika E.",
    title = "{A new characterization of the holographic entropy cone}",
    eprint = "2508.21823",
    archivePrefix = "arXiv",
    primaryClass = "hep-th",
    doi = "10.21468/SciPostPhys.20.4.122",
    journal = "SciPost Phys.",
    volume = "20",
    number = "4",
    pages = "122",
    year = "2026"
}

@article{Ryu:06b20v,
    author = "Ryu, Shinsei and Takayanagi, Tadashi",
    title = "{Holographic derivation of entanglement entropy from AdS/CFT}",
    eprint = "hep-th/0603001",
    archivePrefix = "arXiv",
    reportNumber = "NSF-KITP-06-11",
    doi = "10.1103/PhysRevLett.96.181602",
    journal = "Phys. Rev. Lett.",
    volume = "96",
    pages = "181602",
    year = "2006"
}

@article{Ryu:2006ef,
    author = "Ryu, Shinsei and Takayanagi, Tadashi",
    title = "{Aspects of Holographic Entanglement Entropy}",
    eprint = "hep-th/0605073",
    archivePrefix = "arXiv",
    reportNumber = "NSF-KITP-06-31, KUNS-2021",
    doi = "10.1088/1126-6708/2006/08/045",
    journal = "JHEP",
    volume = "08",
    pages = "045",
    year = "2006"
}

@article{Hubeny:2007xt,
    author = "Hubeny, Veronika E. and Rangamani, Mukund and Takayanagi, Tadashi",
    title = "{A Covariant holographic entanglement entropy proposal}",
    eprint = "0705.0016",
    archivePrefix = "arXiv",
    primaryClass = "hep-th",
    reportNumber = "DCPT-07-13, KUNS-2069",
    doi = "10.1088/1126-6708/2007/07/062",
    journal = "JHEP",
    volume = "07",
    pages = "062",
    year = "2007"
}

@article{Bao:2015bfa,
    author = "Bao, Ning and Nezami, Sepehr and Ooguri, Hirosi and Stoica, Bogdan and Sully, James and Walter, Michael",
    title = "{The Holographic Entropy Cone}",
    eprint = "1505.07839",
    archivePrefix = "arXiv",
    primaryClass = "hep-th",
    reportNumber = "CALT-TH-2015-020, IPMU15-0074, SLAC-PUB-16294, SU-ITP-15-08",
    doi = "10.1007/JHEP09(2015)130",
    journal = "JHEP",
    volume = "09",
    pages = "130",
    year = "2015"
}

@article{Hayden:2011ag,
    author = "Hayden, Patrick and Headrick, Matthew and Maloney, Alexander",
    title = "{Holographic Mutual Information is Monogamous}",
    eprint = "1107.2940",
    archivePrefix = "arXiv",
    primaryClass = "hep-th",
    reportNumber = "BRX-TH-638",
    doi = "10.1103/PhysRevD.87.046003",
    journal = "Phys. Rev. D",
    volume = "87",
    number = "4",
    pages = "046003",
    year = "2013"
}

@article{Hernandez-Cuenca:2023iqh,
    author = "Hern\'andez-Cuenca, Sergio and Hubeny, Veronika E. and Jia, Hewei Frederic",
    title = "{Holographic entropy inequalities and multipartite entanglement}",
    eprint = "2309.06296",
    archivePrefix = "arXiv",
    primaryClass = "hep-th",
    reportNumber = "MIT-CTP/5610",
    doi = "10.1007/JHEP08(2024)238",
    journal = "JHEP",
    volume = "08",
    pages = "238",
    year = "2024"
}

@article{HernandezCuenca:2019wgh,
    author = "Hern\'andez Cuenca, Sergio",
    title = "{Holographic entropy cone for five regions}",
    eprint = "1903.09148",
    archivePrefix = "arXiv",
    primaryClass = "hep-th",
    doi = "10.1103/PhysRevD.100.026004",
    journal = "Phys. Rev. D",
    volume = "100",
    number = "2",
    pages = "026004",
    year = "2019"
}

@article{Czech:2024rco,
    author = "Czech, Bartlomiej and Liu, Yu and Yu, Bo",
    title = "{Two infinite families of facets of the holographic entropy cone}",
    eprint = "2401.13029",
    archivePrefix = "arXiv",
    primaryClass = "hep-th",
    doi = "10.21468/SciPostPhys.17.3.084",
    journal = "SciPost Phys.",
    volume = "17",
    number = "3",
    pages = "084",
    year = "2024"
}

@article{hecdatacurrent,
	author = "Hern\'andez-Cuenca, Sergio",
	title = {{Holographic Entropy Cone Database}},
	year = "2024",
    journal = "\href{https://github.com/SergioHC95/Holographic-Entropy-Cone}{GitHub Repository}"
}

@article{Iizuka:2024die,
    author = "Iizuka, Norihiro and Mukherjee, Arkaprava and Sake, Sunil Kumar and Zenoni, Nicol{\`o}",
    title = "{Multipartite information in sparse SYK models}",
    eprint = "2409.16641",
    archivePrefix = "arXiv",
    primaryClass = "hep-th",
    reportNumber = "YITP-24-118",
    doi = "10.1007/JHEP04(2025)194",
    journal = "JHEP",
    volume = "04",
    pages = "194",
    year = "2025"
}

@article{Kitaev:2005dm,
    author = "Kitaev, Alexei and Preskill, John",
    title = "{Topological entanglement entropy}",
    eprint = "hep-th/0510092",
    archivePrefix = "arXiv",
    reportNumber = "CALT-68-2578, CALT-68-2578",
    doi = "10.1103/PhysRevLett.96.110404",
    journal = "Phys. Rev. Lett.",
    volume = "96",
    pages = "110404",
    year = "2006"
}

@article{Hartman:2013mia,
    author = "Hartman, Thomas",
    title = "{Entanglement Entropy at Large Central Charge}",
    eprint = "1303.6955",
    archivePrefix = "arXiv",
    primaryClass = "hep-th",
    month = "3",
    year = "2013"
}

@article{Levin_2006,
   title={Detecting Topological Order in a Ground State Wave Function},
   volume={96},
   ISSN={1079-7114},
   url={http://dx.doi.org/10.1103/PhysRevLett.96.110405},
   DOI={10.1103/physrevlett.96.110405},
   number={11},
   journal={Physical Review Letters},
   publisher={American Physical Society (APS)},
   author={Levin, Michael and Wen, Xiao-Gang},
   year={2006},
   month=Mar }
\end{document}